\documentclass{elsart}
\usepackage{natbib,epsfig,here}
\begin{document}
\runauthor{J. Schmoll, C.M.Dubbeldam, D.J.Robertson, J.Yao}
\begin{frontmatter}
\title{The Durham Micro-Optics Programme}
\author[Durham]{J.Schmoll,C.M.Dubbeldam,D.J.Robertson,J.Yao}
\footnote{Author contact: jurgen.schmoll@durham.ac.uk, Tel. 0044-(0)191-334-4810}
\address[Durham]{Center for Advanced Instrumentation (CfAI), Netpark Research Institute,Joseph Swan Road, Netpark, Sedgefield TS21 3FB, United Kingdom}

\begin{abstract}

The Durham Microoptics Programme was established to develop key components to be used for integral field spectrographs for upcoming instrumentation projects, focussing on currently existing telescopes as well as on the next generation of ELTs. These activities include monolithic multi-optics machining and grinding, optical surface improvement using various post polishing techniques and replication of micro-optical components. While these developments have mostly slicer-type IFUs in mind, also new types of microlens arrays are in development for fiber based high contrast IFU systems.

\end{abstract}
\begin{keyword}
PACS: 95.55.-n Astronomical and space-research instrumentation; 
42.79.Bh Lenses, prisms and mirrors;
42.79.-e Optical elements, devices, and systems;
42.30.-d Imaging and optical processing; \\
spectroscopy, integral field, 3D-spectroscopy, 2D-spectroscopy, microoptics, image slicer, microlens arrays
\end{keyword}
\end{frontmatter}

\section{Introduction}
The Durham Microoptics Programme is a technology development effort to produce key components for future designs of Integral Field Units (IFUs). The programme has two main threads, image slicer optics and fiber-lenslet systems developing improved performance components particular with respect to surface accuracy and finish, as well as adding novel features such as high contrast microlens arrays. While there is no direct link to any instrumentation project at this stage, this development of new techniques will enable solutions for future projects as KMOS, NIRSPEC IFU or MOMSI.
\section{Image slicer optics}
Over the past few years image slicing IFUs have become increasingly popular, and in many applicatins they are a preferable choice to lenslet or fiber-lenslet designs. Advantages over the fiber-lenslet IFUs include higher efficiency, compact size and a high degree of automatization in manufacture that enables much larger numbers of spatial elements than fiber-lenslet IFUs. Problems to tackle deal with the fact that slicer IFUs are difficult in the context of small reflective surfaces of complicated shape being next to each other. While a sandwich of optically polished surfaces suffers from misalignment and error buildup, the monolithic approach means the elements cannot be produced by classical optical manufacture. After the successful commissioning of the GNIRS IFU in 2003 \cite{All2004}, carrying mirror arrays machined into a monolithic aluminium block, the efforts are now focused in optimizing the manufacture to get smoother surfaces without losing the advantage of machine controlled surface generation. Efforts are made on metal and glass pieces to explore the capabilities of the different materials. As shown in table \ref{methodscompare} \cite{Schi2001, Sto1988}, the choice of the best surface threatment depends not only on material and handling issues, but on the required removal rate and the surface roughness needed.

\begin{table*}[H]
\caption{Removal rates and final surface roughness of different surface generation methods.}
\label{methodscompare}
\begin{center}
\begin{tabular}{l l l}
\hline
Process & Material removal & Typical roughness \\
 & rate [$\frac{mm^3}{s}$] & [nm rms] \\
\hline
Fixed abrasive grinding & 10 & $10^3$ \\
Single point diamond machining & 10 .. 50 & 10..$10^2$ \\
Magneto-Rheological polishing & 0.001 & 3..5 \\
Ion Beam figuring & 0.003 & 0.1..1 \\
\hline
\end{tabular}
\end{center}
\end{table*}

\subsection{Metal micro optics}
The effort in metal optics generation focuses on two different techniques. At first, the precision of diamond machining using ultra-high precision (UHP) machines depends on various parameters such as machine setup, tool type and material being machined. The CfAI uses a Nanotech 350FG 5-axis UHP machine to create free form surfaces to nanometer precision. Suitable metrology equipment allows the quick measurement of the workpieces roughness (ZYGO white light interferometer) and shape (non contacting profilometer, FISBA phase shifting interferometer). The second technique is the replication of metal image slicer optics, being studied within the EU FP6 OPTICON {\it Smart Focal Planes} network (see {\it Optical replication techniques for image slicers} within these proceedings).

\subsection{Glass micro optics}
Due to the different properties of glass when being ground or polished, tests have also been started on glass optics. One goal is the construction of a full working glass slicer IFU, using monolithic optic arrays for the pupil and slit mirrors. Apart from this, the roughness improvement of ground free-form surfaces is subject to tests of post-polishing techniques to remove the residuals of the grinding process. 

\subsubsection{Partly monolithic architecture:} 
\begin{center}
\begin{figure}[H]
\epsfig{file=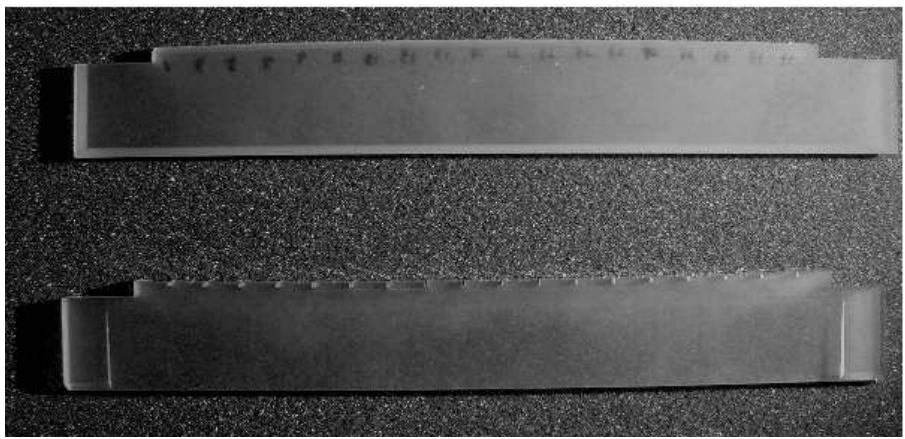, width=120mm}
\caption{Pupil (below) and slit (above) mirror arrays of the prototype glass IFU.}
\label{glassarrays}
\end{figure}
\end{center}

\begin{center}
\begin{figure}[H]
\epsfig{file=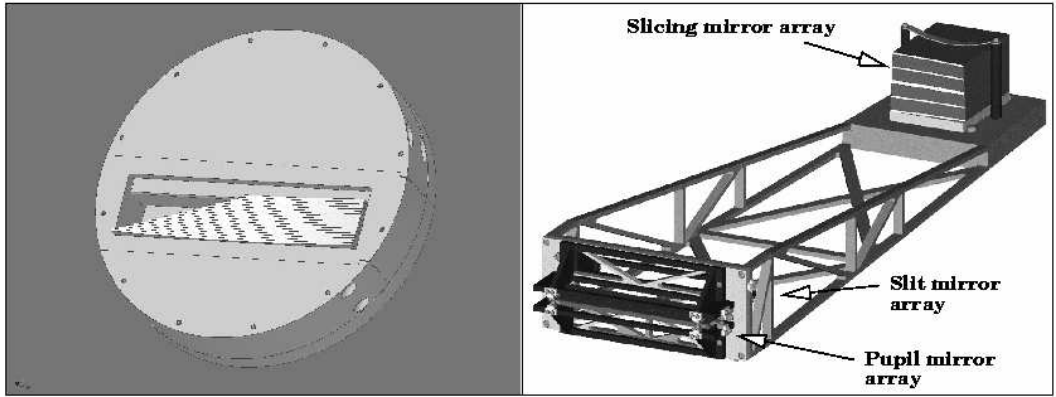, width=140mm}
\caption{Schematic slicing mirror assembly (left) and complete prototype concept.}
\label{protoslicer}
\end{figure}
\end{center}

This glass slicer IFU prototype consists of monolithic Zerodur parts for the pupil and slit mirror arrays (see fig. \ref{glassarrays}), produced by free-form grinding. The slicing mirror array is made from single blanks to maintain a fill factor near to 100 \%. The stack of blanks is assembled to a jig (see fig. \ref{protoslicer}, left), allowing conventional grinding and polishing of a spherical surface before sliding the different slices into final alignment. Critical issues are the precision of the ground datum surfaces of the blanks, their thickness and the flatness of the corresponding, diamond machined jig surfaces. In their final position the slices are held in place by clamp tension of the jig that becomes part of the IFU.

\subsubsection{Surface smoothening:}
This approach focuses on post-polishing processes, reducing surface roughness while maintaining the optical figure. The goal is to treat complete arrays after the optical surfaces have been ground to their nominal optical shape. For the experiments, flat Schott Zerodur test parts have been diamond ground to roughness values of 10 to 20 nm rms. Three different smoothing techniques are under investigation:
\begin{description}
\item{\bf Ion-beam etching:} After spin-coating the surface using $SiO_2$, it is treated in an ion beam that ablates the coating with the same rate as the substrate. Hence, the smoother surface of the coating is transferred into the substrate \cite{Schi2001}. Initial tests by IOM\footnote{IOM Institute for Surface Modification, Leipzig, Germany} revealed that the roughness could be reduced from 20..25 nm rms to about 4 nm rms on small sampling areas, albeit the diamond tool marks are still visible using an atomic force microscope. 
\item{\bf Liquid jet polishing:}
Developed and applied by TNO\footnote{TNO Delft, The Netherlands}, a controlled jet of polishing liquid removes surface irregularities. Currently TNO uses a converted polishing machine to test this technique on the Zerodur test pieces.
\item{\bf Magneto-rheological jet polishing (MR-JET):}
Another type of liquid jet polishing where the abrasive particles are oriented by means of a magnetic field that also allows a better beam control \cite{kor2004}. Tests have been done by QED\footnote{QED Technologies, Rochester NY., USA}, indicating a roughness improvement by about a factor of two.
\end{description}
A side-by-side comparision of these techniques will allow a determination of the most suitable one for smoothening out the residual roughness produced by the diamond grinding. These techniques will possibly allow an expansion of the IFU application to visual wavelength using partly monolithic slicer hardware.   

\section{New microlens array developments}
\begin{center}
\begin{figure}[H]
\epsfig{file=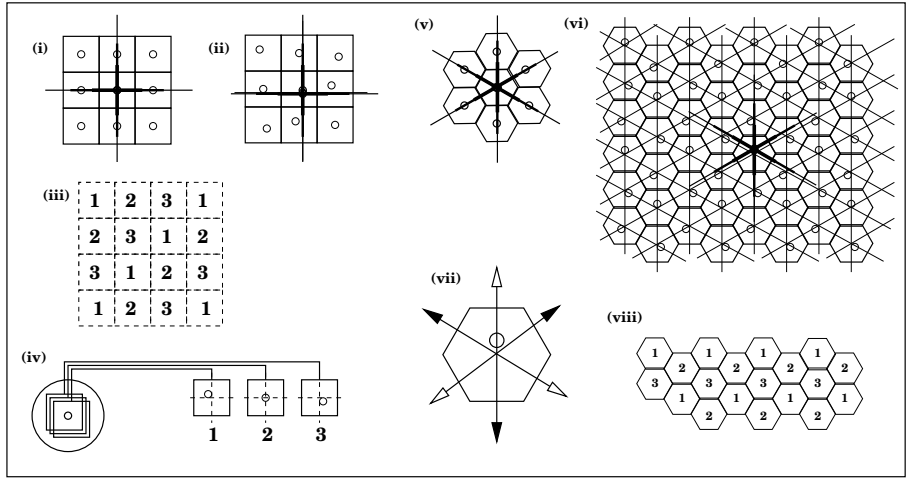, width=120mm}
\caption{Dithered lenslet arrays: (i) Square, standard. (ii) Square, dithered. (iii) Tiling pattern. (iv) Lenslet types. (v) to (viii) similar for hexagonal arrays.}
\label{larr_dither}
\end{figure}
\end{center}

The CfAI has designed and built several fiber-lenslet IFUs, amongst them recently two for the GMOS spectrographs for both Gemini telescopes \cite{All2002} and the 2000 element IFU for the IMACS spectrograph of the Magellan-I telescope \cite{Sch2004}. In all cases during construction the performance of the epoxy-on-glass microlens arrays used was degraded by scatter, hence a selection had to be made by purchase of more arrays than actually needed. The roughness of the lens surfaces was between 40 and 210 nm rms, hence easily visible in a microscope and causing significant light scatter. Using the CfAI UHP machine, the tooling for new microlens arrays can now be done in-house, reducing costs dramatically and allowing variations of the lenslet types and materials. To improve the quality of microlens arrays, tool production is planned for epoxy on glass arrays as well as for monolithic glass molding processes. This will allow a tradeoff between both types of microlens architecture. In addition the monolithic glass approach allows a larger wavelength coverage, and a higher temperature stability is expected, being important for cryogenic use.\\
Another problem of fiber-lenslet coupling is the possible cross-talk between adjacent elements, caused by borderline effects and diffraction. If one element is illuminated by a bright source, this will cause spikes in four (see \cite{Rot2005}) or six directions, depending on the lenslet shape whether square or hexagonal. In a regular fiber lenslet coupling pattern this light can enter adjacent fibers. Altough the effect is small, it becomes critical where dim sources next to bright sources are to be observed, as in spectroscopy of QSO host galaxies, circumstellar disks or possibly extrasolar planets. To avoid this while maintaining the fill factor and regular spatial sampling, so-called dithered lenslet arrays have been designed to assure that a spike will miss the first and second neighbour on each side (fig. \ref{larr_dither}). A suitable fiber-lenslet coupling causes the fiber to act as a filter, rejecting spikes coming from the third neighbour and beyond because the angle of incidence is beyond the numerical aperture of the fiber. This condition is met if equation (\ref{filterequation}) is fulfilled: The numerical aperture (NA) selection depends on the microlens focal ratio N and the refractive index $n_0$ outside the fiber. For example, using f/10-lenslets in a monolithic block of $n_0=1.5$ with the fibers index-matched to it, the maximum NA is 0.36. As the most commonly used step-index fibers have NA values about 0.22, the filter will work for a wide range of applications.

\begin{equation}
NA \leq n_0 \sin \left( \arctan \left( \frac{2.5}{N} \right) \right) \
\label{filterequation}
\end{equation}

As shown in fig. \ref{larr_dither}, three different lenslet types are required. For square lenslets there are one concentric and two similar off axis lenses turned by 180$^\circ$ with respect to each other. In the hexagon case, one off-axis type is needed for three subsequent orientations, differing in rotation angle by 120$^\circ$. Using this tiling method, in every spike direction the next two neighbours do not see the spikes. Manufacture of monolihic arrays of these types is possible using freeform generators like the CfAI UHP machine mentioned above. The production of normal and dithered lenslet arrays by the slow tool servo technique has been simulated successfully. The next step will be the manufacture of lenslet array masters being used for epoxy-on-glass types as well as for monolithic glass arrays. 

\begin{thebibliography}{999}
\bibitem{All2002} Allington-Smith, J.R. et al. 2002, PASP Volume 114, Issue 798, pp. 892-912
\bibitem{All2004} Allington-Smith, J.R. et al. 2004, SPIE Proc Volume 5492, pp. 701-710
\bibitem{kor2004} Kordonski, William; Shorey, Aric B.; Sekeres, Arpad; 2004, SPIE, Volume 5180, pp. 107-114
\bibitem{Rot2005} Roth, M.M. et al. 2005, PASP Volume 117, Issue 832, pp. 620-642
\bibitem{Schi2001} Schindler, Axel et al. 2001, Proc. SPIE Vol. 4440, p. 217-227
\bibitem{Sch2004} Schmoll, J. et al. 2004, SPIE Proc, Volume 5492, pp. 624-633
\bibitem{Sto1988} Stowers, I.F., Komanduri, R., Baird, E.D. 1988, SPIE Proc Volume 966, pp. 66-73
\end{thebibliography}
\end{document}